\documentclass[aps,prl,reprint,superscriptaddress,floatfix]{revtex4-1}

\usepackage{graphicx}
\usepackage{placeins}

\begin{document}


\title{Two-photon quantum interference from separate nitrogen vacancy centers in diamond}


\author{Hannes Bernien}
\email{h.bernien@tudelft.nl}
\affiliation{\nolinebreak[4]{Kavli Institute of Nanoscience Delft, Delft University of Technology, PO Box 5046, 2600 GA Delft, The Netherlands}}
\author{Lilian Childress}
\affiliation{Department of Physics and Astronomy, Bates College, 44 Campus Avenue, Lewiston, Maine 04240, USA}
\author{Lucio Robledo}
\affiliation{Kavli Institute of Nanoscience Delft, Delft University of Technology, PO Box 5046, 2600 GA Delft, The Netherlands}
\author{Matthew Markham}
\affiliation{Element Six, Ltd., Kings Ride Park, Ascot, Berkshire SL5 8BP, United Kingdom}
\author{Daniel Twitchen}
\affiliation{Element Six, Ltd., Kings Ride Park, Ascot, Berkshire SL5 8BP, United Kingdom}
\author{Ronald Hanson}
\affiliation{Kavli Institute of Nanoscience Delft, Delft University of Technology, PO Box 5046, 2600 GA Delft, The Netherlands}

\date{\today}

\begin{abstract}

We report on the observation of quantum interference of the emission from two separate nitrogen vacancy (NV) centers in diamond. Taking advantage of optically induced spin polarization in combination with polarization filtering, we isolate a single transition within the zero-phonon line of the nonresonantly excited NV centers. The time-resolved two-photon interference contrast of this filtered emission reaches 66\%. Furthermore, we observe quantum interference from dissimilar NV centers tuned into resonance through the dc Stark effect. These results pave the way towards measurement-based entanglement between remote NV centers and the realization of quantum networks with solid-state spins.

\end{abstract}


\maketitle

The nitrogen vacancy (NV) center in diamond is an attractive candidate for the construction of quantum networks~\cite{Briegel1998, Kimble2008, Childress2006a}.
It is a highly stable single photon emitter \cite{Kurtsiefer2000}, whose optical transitions are linked \cite{Togan2010} to a long-lived electronic spin with excellent coherence properties \cite{Balasubramanian2009}. 
The electronic spin can be coherently manipulated with high fidelity with pulsed microwave fields~\cite{Jelezko2004, deLange2010}. By extending these techniques to proximal nuclear spins, a controllable multiqubit quantum register can be realized \cite{Dutt2007, Neumann2008, Robledo2011}.

Optical schemes based on photon interference provide a way to establish quantum interactions over a distance. When two photons that are indistinguishable simultaneously impinge on a beamsplitter they coalesce into the same output port~\cite{Hong1987}. If these two photons are each entangled with the spin of the emitter, their interference can be exploited to generate entanglement between two distant emitters~\cite{Duan2001, Simon2003, Barrett2005}. This type of measurement-based entanglement has recently been achieved between two remote single ions~\cite{Moehring2007}. While spin-photon entanglement has been realized using NV centers~\cite{Togan2010}, two-photon quantum interference (TPQI) with NV centers has so far not been observed. 


Here we capitalize on recent advances in the understanding of the NV center's excited states~\cite{MazeDoherty2011,Batalov2009,Kaiser2009,Fu2009,Bassett2011} and a drastic improvement of the photon collection efficiency~\cite{Hadden2010, Robledo2011}, and demonstrate TPQI from two separate NV centers. Furthermore, we show that even dissimilar NV centers can exhibit TPQI when their emission lines are brought into resonance using  dc Stark tuning~\cite{Tamarat2006, Bassett2011}. This latter technique greatly facilitates a  scaling to larger networks.

\begin{figure}
\includegraphics[width=\linewidth]{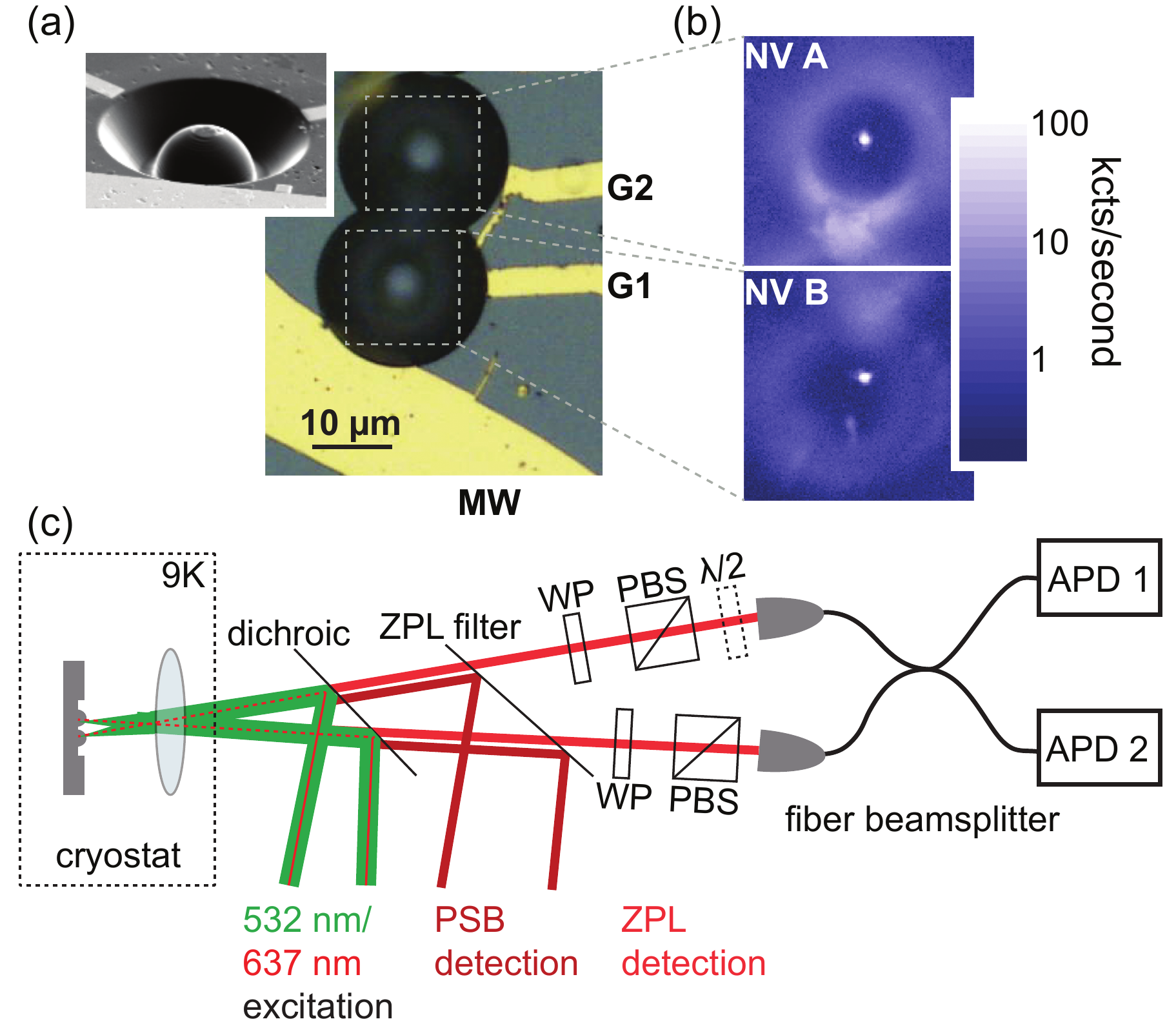}
\caption{\label{sample}(a) Optical microscope image of the sample. The microwave stripline (MW) is used for spin manipulation and the gates (G1 and G2) are used for Stark tuning of the optical transition energies. Inset: Scanning electron microscope image of a similar device. (b) Confocal microscope image showing NV A and NV B (logarithmic color scale). (c) Experimental setup. The two NV centers inside the same diamond are simultaneously excited with either a resonant (637$\,$nm) or an nonresonant (532$\,$nm) laser. Two separate paths allow detection of photons emitted into the zero-phonon line (ZPL) or the phonon side band (PSB). A variable retardance waveplate (WP) aligns the polarization of the desired transition to the polarizing beamsplitter (PBS) while compensating for ellipticity introduced by other optical components. }
\end{figure}

We perform our experiments on naturally occurring NV centers in high purity type IIa chemical-vapor deposition grown diamond~\cite{Isberg2002} with a $\langle$111$\rangle$ crystal orientation obtained by cleaving a $\langle$100$\rangle$ substrate.  Because a TPQI measurement involves coincidence detection of photons emitted from two centers, high collection efficiency is essential. To this end, we deterministically fabricate solid immersion lenses~\cite{Hadden2010, Siyushev2010} around preselected centers by focused ion beam milling. Figure~1(a) shows an optical microscope image of the device being used; the inset is an scanning electron microscope image of a similar device.  In the confocal scans [Fig.~1(b)], the NV centers  appear as bright spots inside the lenses. To enable spin manipulation and Stark tuning of the optical transition energies, we lithographically define a gold microwave stripline and gates around the solid immersion lenses. A dual path confocal setup allows us to individually address two NV centers within the same diamond  [Fig.~1(c)]. The sample is mounted inside a flow cryostat and experiments are performed at a temperature of 9$\,$K.

The optical emission spectrum of the NV center [Fig.~2(a)] consists of both direct transitions between the ground and the excited state [the so-called zero phonon line, ZPL, that contributes 4\% to the total emission] and transitions that additionally involve the emission of phonons [phonon side band (PSB)]. At low temperatures, the ZPL emission spectrum exhibits several narrow lines which, for low-strain centers, correspond to spin selective transitions between the ground and excited state~\cite{MazeDoherty2011}. Observation of TPQI requires indistinguishable photons which we produce by isolating a single transition within the ZPL. Appropriate bandpass filters remove the incoherent fraction of the emission (PSB). Nonresonant excitation, as used in this experiment, polarizes the NV electronic spin into the $m_s=0$ state, hence only transitions between the  $E_x$ and  $E_y$ excited states ($m_s=0$) and the  $m_s=0$ ground state level occur. The dipoles associated with these two transitions are orthogonal to each other and to the N-V axis; by working with NV centers oriented along the $\langle 111 \rangle$ direction we ensure that collected ZPL photons remain linearly polarized, with orthogonal polarizations for $E_x$ and $E_y$~\cite{Fu2009, Kaiser2009}. 
Consequently, for a center that is spin polarized into $m_s=0$, we can isolate the $E_x$ or $E_y$ emission line by placing a polarizer in the detection path. Furthermore, we maximize the signal by using the polarization selectivity to excite one dipole predominantly.

To characterize the ZPL fine structure, we use photoluminescence excitation (PLE) to probe the ZPL absorption spectrum and a scanning Fabry-Perot (FP) cavity to analyze the ZPL emission spectrum. PLE spectra are obtained by sweeping the frequency of the resonant excitation laser [637$\,$nm, Fig.~1(c)] across the ZPL transitions while recording the red-shifted emission into the PSB. During PLE scans, we ensure the correct charge state and visibility of the $m_s=0$ transitions by simultaneous weak green excitation \cite{Jelezko2002}. Figure~2(b) shows the PLE spectra of the two selected emitters. While the lower-energy $E_y$ transitions are 2.9$\,$GHz apart from each other, the higher-energy $E_x$ transitions partly overlap. This near overlap is also observed by using a scanning FP cavity to monitor the ZPL emission under nonresonant (532$\,$nm) excitation [Fig.~2(c)]. By setting the filters to transmit only the linearly polarized light associated with the $E_x$ transitions, we observe a single emission line from each NV center with nearly equal frequencies.

Inhomogeneous broadening of the emission linewidths will limit the observability of quantum interference~\cite{Legero2003}. For individual PLE scans, obtained in the absence of green light, we find narrow 38$\pm7\,$MHz ($36\pm5\,$MHz) linewidths for NV A (NV B); however, a 532$\,$nm repump pulse between scans leads to an overall distribution of frequencies with an inhomogeneous linewidth of $263\pm6\,$MHz for NV A and $483\pm5\,$MHz for NV B~\cite{Fu2009, Robledo2010}. The same spectral diffusion is observed in PLE spectra recorded with simultaneous green excitation [Fig.~2(b)] and in the ZPL emission spectrum under 532$\,$nm excitation [Fig.~2(c)]. Although the observed inhomogeneous broadening exceeds the radiative linewidth by an order of magnitude, two-photon interference effects can still be detected. Provided that the emission linewidth does not exceed the inverse time resolution of the photon detectors, simultaneous detection of a photon from each NV center erases the which-path frequency information, allowing quantum interference to be observed~\cite{Legero2004, Lettow2010}.

 \begin{figure}
 \includegraphics[width=\linewidth]{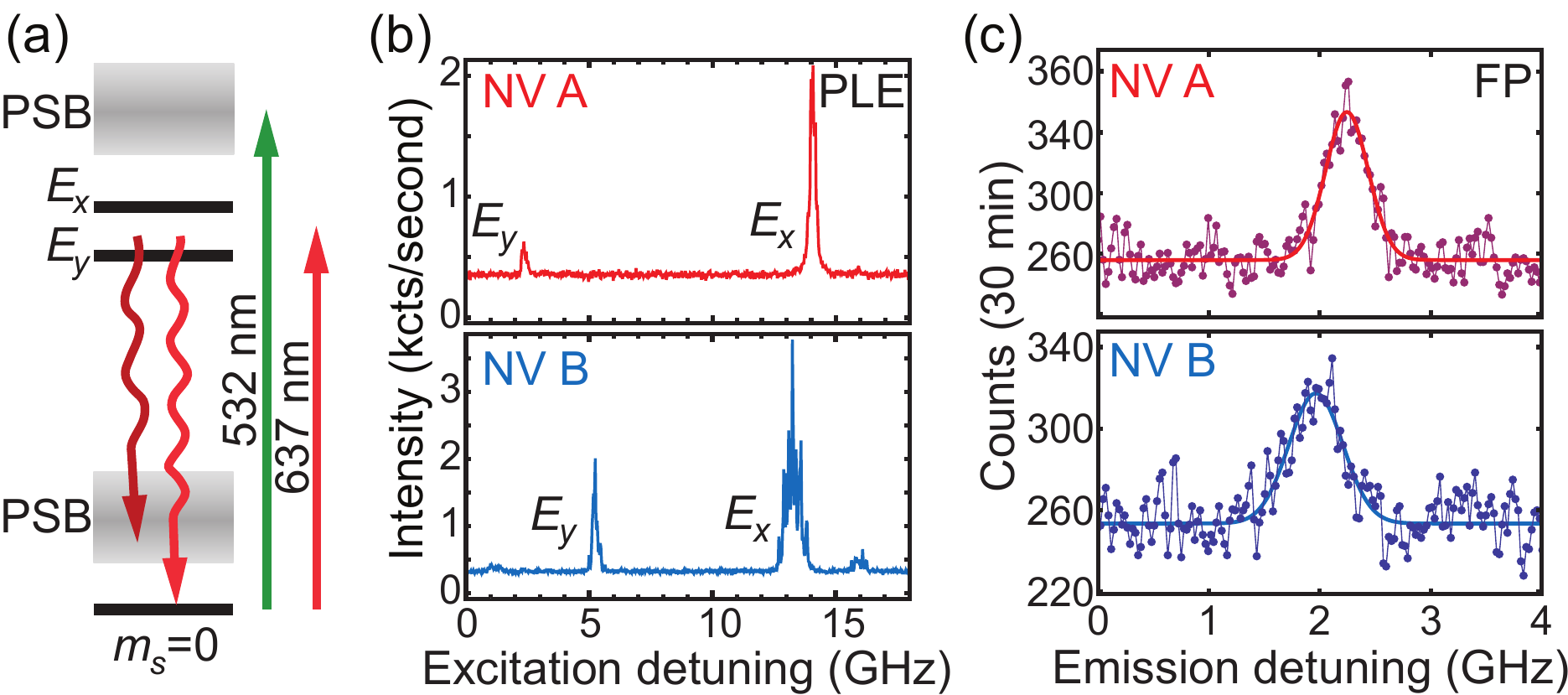}
 \caption{\label{spectra}Spectral characterization of NV A and NV B. (a) Level structure. The $m_s=0$ ground state is connected via spin selective transitions to the $E_x$ and $E_y$ ($m_s=0$) excited state, $m_s=\pm1$ states are omitted for clarity. Photons that are emitted into the phonon side band (PSB) can be spectrally separated from resonant emission. Also shown are the resonant (637$\,$nm) and the nonresonant laser (532$\,$nm). (b) Resonant excitation of the ZPL and detection of the PSB. The frequency of the red laser is swept across the resonances while weak green excitation prevents photoionization and optical pumping (see text). The spectra are averaged over 70 scans. (c) Scanning Fabry-Perot spectrum under nonresonant excitation. Only the $E_x$ transition is visible confirming sufficient suppression of the $E_y$ transition by polarization filtering. The linewidths (450$\,$MHz for NV A and 570$\,$MHz for NV B) include drifts of the cavity during the measurement. The offset is due to detector dark counts.}
 \end{figure}

For the interference measurement, we employ a green (532$\,$nm) pulsed laser (62$\,$ps) with a repetition frequency of 10$\,$MHz both to excite the two NV centers and to initialize the spin state into $m_s=0$. A combination of spectral filtering (ZPL filter) and polarization rejection (as discussed above) isolates the $E_x$ lines of NV A and B.  This emission is then coupled into the two input ports of a polarization-maintaining fiber beamsplitter, ensuring excellent spatial mode overlap. We establish temporal mode overlap by using equidistant excitation and collection paths for the two centers. The output ports of the beam splitter are connected to two avalanche photo diodes (APDs) with sub-ns time resolution; the APD signals trigger the start and the stop of a fast counting module whose jitter is less than 12$\,$ps. By recording the coincidence counts as a function of start-stop delay, we perform a time-resolved measurement.  

As a calibration experiment, we insert a $\lambda/2$ wave plate into one detection path [Fig.~1(c)] so that the two photons entering the beamsplitter have orthogonal polarization. This makes the two photons distinguishable and no interference can be observed. Therefore, the coincidence distribution only reveals the temporal overlap of two independent photon wave packets [see Fig.~3(a)]. 
The situation changes dramatically when the photons enter the beamsplitter with parallel polarization [Fig.~3(b)]. For this case, two-photon quantum interference is observed: around zero detection time difference ($\tau=0$) the two photons mainly leave the beamsplitter into the same output port, leading to a significant reduction in coincidence detections. For larger time differences interference is concealed because of averaging over many photons with different frequencies~\cite{Legero2003}.  We thereby observe TPQI as a reduction in coincidence detection events within a time window given by the inverse of the inhomogeneous emission linewidth.  

In principle, full visibility interference can be observed even for inhomogeneously broadened and detuned photon sources at the cost of a reduced width of the interference dip~\cite{Legero2004}. The observed contrast at $\tau = 0$ is limited primarily by NV emission into undesired spectral lines owing to imperfect control over the charge and spin state of the center. Specifically, when the center is in the neutral charge state (NV$^0$), a portion of its broadband fluorescence lies within the 3$\,$nm bandwidth of our ZPL filters; from independently measured optical spectra, these NV$^0$ photons contribute approximately 10$\%$ to the collected emission.  Furthermore, under 532$\,$nm excitation a residual $\sim$10$\%$  $m_s=\pm1$ spin occupation~\cite{Robledo2011} produces circularly polarized emission on several other transitions, contributing $\sim$5$\%$ of our polarization-filtered signal.  From such a $15\%$ background level we expect at most 72\% visibility. Furthermore, finite time resolution in our detection system will average over sharp temporal features, raising the depth of the zero-time-difference minimum.  

We observe a quantitative agreement between our data and no-free-parameter simulations of the experiment.  Following Legero et al.~\cite{Legero2003}, we model TPQI of exponentially decaying photon wavepackets with Gaussian frequency noise and calculate the expected coincidence detections using independently measured parameters.  At $\tau=0$ we observe
 a contrast of 66$\pm$10\% which is to our knowledge the highest value reported for two separate solid-state emitters~\cite{Sanaka2009, Patel2010, Flagg2010, Lettow2010}. This value can be improved by more stringent filtering of the ZPL emission or by increased control over the spin and charge states. We note that for measurement-based entanglement, the visibility determines the fidelity while the width of the interference dip sets the success probability of the entanglement operation.

 \begin{figure}
 \includegraphics[width=\linewidth]{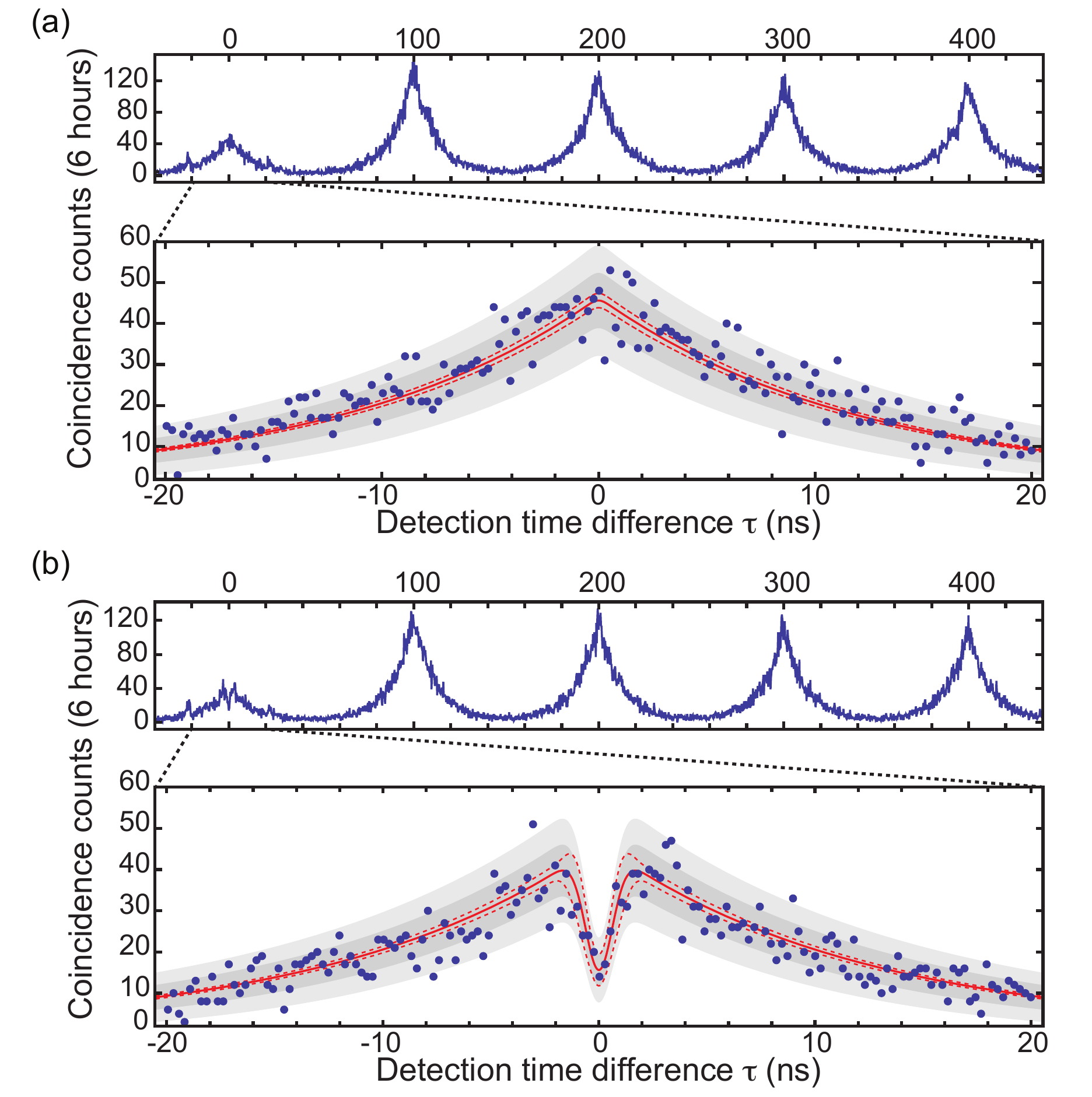}
 \caption{\label{tpqi}Two-photon quantum interference. (a)~Orthogonal polarization. Periodic peaks of the coincidence counts correspond to the 10$\,$MHz repetition frequency of the excitation. The coincidence distribution shows typical bunching and antibunching features of the two independent NV centers under pulsed excitation~\cite{Beveratos2002}. (b) Parallel polarization. Interference of indistinguishable photons leads to a significant decrease in coincidence events at zero time delay. Simulations (solid line) using only independently measured parameters are done according to Legero et al.~\cite{Legero2003}. The dashed lines show 1 standard error uncertainty in the simulation. The dark and light grey areas illustrate 1 and 2 standard deviations expected scatter. Parameters used for the simulations are: excited state lifetime = 12.0$\pm0.4\,$ns, detuning between the two centers~=~130$\pm150\,$MHz, frequency jitter between them~=~550$\pm80\,$MHz, APD jitter~=~410$\pm10\,$ps, dark count rate~=~60$\pm10\,$s$^{-1}$, NV~B count rate~=~1470$\pm50\,$s$^{-1}$, NV~A count rate~=~2700$\pm50\,$s$^{-1}$, background~=~15$\pm5\,$\%. 
}
 \end{figure}

To observe TPQI, we selected NV centers with nearly identical $E_x$ frequencies;  in general, however, because of their different strain environments, two NV centers are unlikely to exhibit the same emission frequencies~\cite{Manson2006, Batalov2009}. Even in high purity type IIa samples, we typically observe a spread of tens of GHz between different centers. Nevertheless,  NVs can be tuned into resonance by applying electrical fields to induce dc Stark shifts of the transition energies~\cite{Tamarat2006}. The tuning range of our devices reaches several GHz and is enhanced to tens of GHz in the presence of 532$\,$nm excitation~\cite{Bassett2011}. This enables us to observe TPQI in the general case of dissimilar sources. 

In the absence of an external voltage bias, the $E_y$ transitions of NV A and NV B lie far apart in energy so that their TPQI cannot be measured within the time resolution of our detectors. By applying a voltage to gate~1 [Fig.~1(a)] while keeping the microwave stripline and gate~2 on ground, we can tune the  $E_y$ transitions into resonance. Because the degree of tuning is strongly affected by the presence of nonresonant excitation~\cite{Bassett2011}, we calibrate the Stark tuning under the same pulsed 532$\,$nm excitation conditions as used for the interference measurement: the PLE spectra in Fig.~4(a) show the additional fluorescence induced by resonant excitation, revealing the spectral location of the  $E_y$ line. Near -13.6\,V, we observe overlapping transition energies for the two NV centers, establishing the appropriate setting for observation of TPQI. In addition to shifting the energy of the NV center transitions, applied electric fields can also rotate the axes of the  $E_x$  and $E_y$ dipoles; we find that significantly different polarization settings are required to filter the $E_y$  emission at this gate voltage. 

Using the calibrated voltage and polarization settings, we measure the time-resolved interference of photons emitted on the $E_y$ transitions from each NV center. Figure~4(b) shows a significant decrease in coincidence detection events at zero time difference. The width of the interference signal is smaller than what would be expected from earlier measurements of the inhomogeneous linewidth of the emitters. Because the laser intensity changes the effective Stark shift, a drift in sample position directly translates into a spectral shift. To account for the spectral variations, we perform the measurement by alternating PLE spectroscopy and minute-duration coincidence detection, and post-select the data based on their relative detunings. 
We find our data to agree with a simulation based on the measured frequency distribution [solid line in Fig.~4(b)]. In contrast, a model without TPQI [shaded area in Fig.~4(b)] fails to reproduce the observed drop in coincidences around $\tau = 0$.  
Increased control over spatial and laser power fluctuations is expected to greatly enhance the interference contrast in the presence of gate voltage tuning.


We have demonstrated two-photon quantum interference with separate NV centers in diamond. The observed contrast can be further improved through resonant excitation, which eliminates photon emission from the incorrect charge and spin states. Moreover, the coincidence rates may be enhanced by embedding NV centers into optical cavities \cite{Aharonovich2011, Faaraon2011}. When combined with recently-demonstrated spin-photon entanglement~\cite{Togan2010}, our results enable remote entanglement of NV centers, 
and open the door to applications in quantum information processing and long distance quantum communication.

 \begin{figure}
 \includegraphics[width=\linewidth]{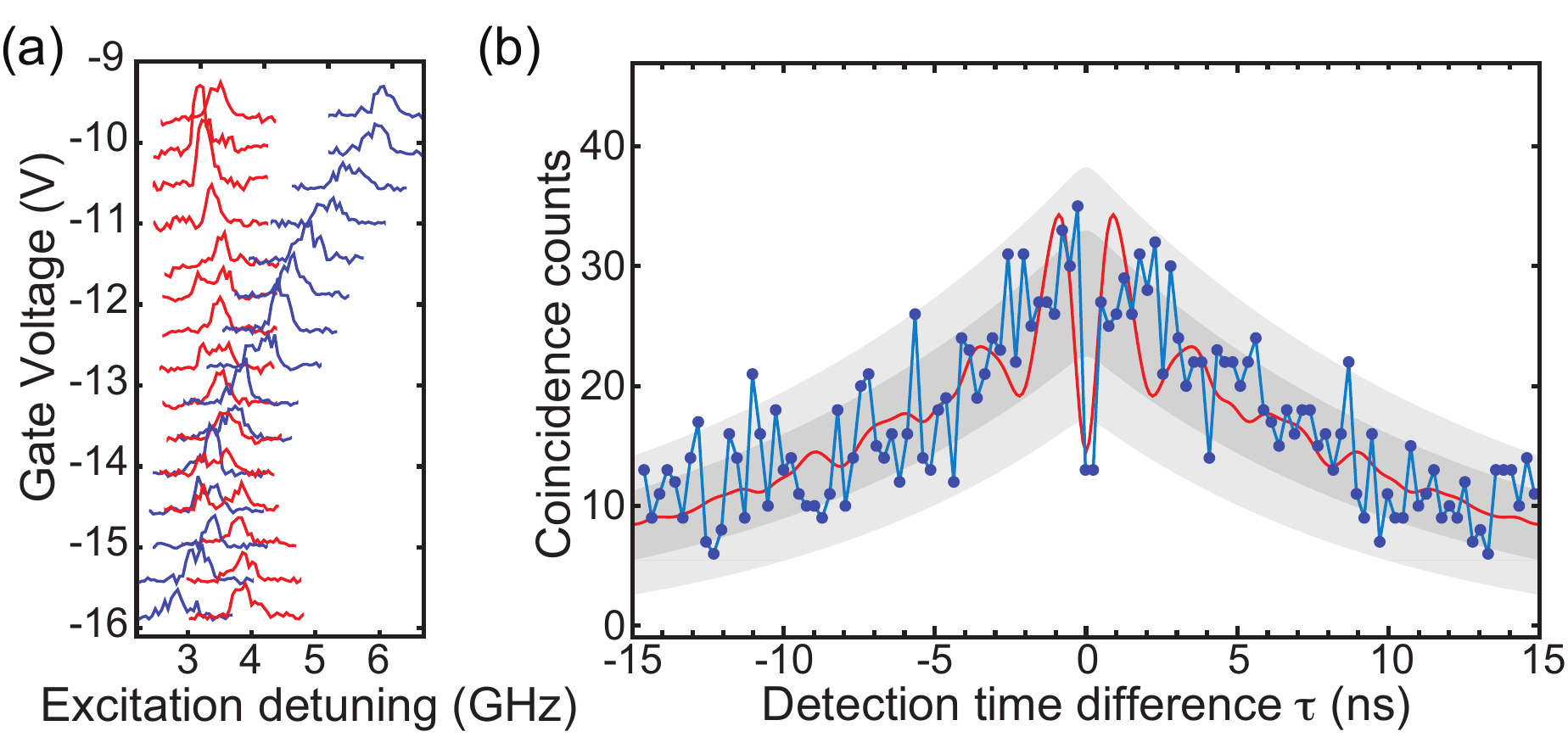}
 \caption{\label{tpqi_tuning}Interference of two dissimilar sources. (a) $E_y$ transition energy as a function of applied gate voltage. The two centers show opposite tuning behavior and are brought into resonance at -13.6V. Data are taken under simultaneous green excitation with the same power that is used for the interference measurement. (b) Two-photon quantum interference. The overall coincidence distribution is the sum of 255 one-minute histograms. Before each interference measurement we perform a PLE scan. We select only histograms for which the relative detuning of the two centers is between 350 and 1200$\,$MHz. The simulation (red line) is based on the measured frequency distribution and shows oscillations that reflect the discrete set of detuning values. The dark and light grey areas illustrate the expected signal for two noninterfering sources with 1 and 2 standard deviations, respectively.}
 \end{figure}

We thank B. Hensen, V. Jacques, M. D. Lukin, Y. Rezus and E. Togan for helpful discussions. This work is supported by the Dutch Organization for Fundamental Research on Matter (FOM), the Netherlands Organization for Scientific Research (NWO), the DARPA QuEST program and the EU DIAMANT and SOLID programs. M. M. and D. T. acknowledge support of the DARPA QuASAR program.


\end{document}